\documentclass[12pt, a4paper]{article}
\usepackage{authblk}
\usepackage{amsmath}
\usepackage{csquotes}
\usepackage{graphicx}
\usepackage[style=authoryear, backend=bibtex, maxcitenames=2,
			      maxbibnames=99, dashed=false,
						sorting = nyt, sortcites = true,
            firstinits=true, uniquename=false]{biblatex}
\usepackage[margin=3cm]{geometry}
\newcommand{\ksi}{\kappa_v^{i*}}
\newcommand{\kis}{\kappa_v^{s*}}
\newcommand{\csi}{\mathcal{C}_v^{i*}}
\newcommand{\cis}{\mathcal{C}_v^{s*}}
\newcommand{\Tsi}{\mathcal{T}_{si}^*}
\newcommand{\Tis}{\mathcal{T}_{is}^*}
\newcommand{\Csi}{\mathcal{C}_{si}^*}
\newcommand{\Cis}{\mathcal{C}_{is}^*}
\newcommand{\A}{\textbf{A}}

\begin{document}
\title{Hierarchical structures in Northern Hemispheric extratropical winter ocean-atmosphere interactions}
\author[1,2]{Marc Wiedermann\footnote{marcwie@pik-potsdam.de}}
\author[1,3]{Jonathan F. Donges}
\author[4]{D\"orthe Handorf}
\author[1,2,5,6]{J\"urgen Kurths}
\author[1]{Reik V. Donner}
\affil[1]{Potsdam Institute for Climate Impact Research --- P.O. Box 60 12
	03, 14412 Potsdam, Germany, EU}
\affil[2]{Department of Physics, Humboldt University --- Newtonstr. 15,
	12489 Berlin, Germany, EU}
\affil[3]{Stockholm Resilience Centre, Stockholm University --- Kr\"aftriket
	2B, 114 19 Stockholm, Sweden, EU}
\affil[4]{Alfred Wegener Institute, Helmholtz Centre for Polar and Marine
Research --- Telegrafenberg A43, 14473 Potsdam, Germany}
\affil[5]{Institute for Complex Systems and Mathematical Biology,
  University of Aberdeen --- Aberdeen AB24 3FX, UK, EU}
\affil[6]{Department of Control Theory, Nizhny Novgorod State University ---
}
\maketitle
\begin{abstract}
In recent years extensive studies on the Earth's climate system have been
carried out by means of advanced complex network statistics. The great majority
of these studies, however, have been focusing on investigating correlation
structures within single climatic fields directly on or parallel to the
Earth's surface. Here, we develop a novel approach of node weighted coupled
network measures to study correlations between ocean and atmosphere in the
Northern Hemisphere extratropics and construct 18 coupled climate networks, each consisting
of two subnetworks. In all cases, one subnetwork represents monthly sea-surface
temperature (SST) anomalies, while the other is based on the monthly
geopotential height (HGT) of isobaric surfaces at different pressure levels
covering the troposphere as well as the lower stratosphere.  The weighted
cross-degree density proves to be consistent with the leading coupled pattern
obtained from maximum covariance analysis. Network measures of higher order
allow for a further analysis of the correlation structure between the two
fields and consistently indicate that in the Northern Hemisphere extratropics
the ocean is correlated with the atmosphere in a hierarchical fashion such that
large areas of the ocean surface correlate with multiple statistically
dissimilar regions in the atmosphere. Ultimately we show that, this observed hierarchy is
linked to large-scale atmospheric variability patterns, such as the Pacific North American
pattern, forcing the ocean on monthly time scales. 
\end{abstract}

\subparagraph*{Keywords:}
coupled climate networks, extratropical ocean-atmosphere
interaction, node-weighted network measures, hierarchical networks

\section{Introduction}
In the last years, complex network analysis has been established as a powerful
tool to study statistical interdependencies in the climate system
\parencite{donges_complex_2009, tsonis_architecture_2004, tsonis_role_2008,
	tsonis_what_2006, donges_how_2015}. Links in the so-called climate networks
represent functional interdependencies indicated by significant correlations
\parencite{donges_backbone_2009, donges_complex_2009, radebach_disentangling_2013,
	palus_discerning_2011} or the synchronous occurrence of extreme events
\parencite{stolbova_topology_2014, boers_complex_2013, malik_spatial_2010,
	malik_analysis_2011, boers_complex_2014} in climatic time series taken at
different grid points or measurement sites on or parallel to the Earth's
surface. 

In addition to studies on observational data of climate dynamics,
climate networks have also been applied successfully to hindcast extreme
events, such as extreme precipitation in South America
\parencite{boers_prediction_2014}, or to predict the occurrence of El Ni\~no
episodes \parencite{ludescher_improved_2013,ludescher_very_2014} and discriminate
between different event types
\parencite{wiedermann_climate_2016, radebach_disentangling_2013,tsonis_topology_2008,yamasaki_climate_2008}.

So far, most studies conducted within the framework of climate networks focused
solely on the dynamics within a single climatic field or layer. Besides
atmospheric characteristics like surface air temperature, sea level pressure,
or precipitation, recent studies have also addressed ocean dynamics represented
by ocean temperature variability at the surface
\parencite{feng_are_2014,tantet_interaction_2014} or different depths
\parencite{van_der_mheen_interaction_2013} as well as the spatio-temporal
variability in the strength of the Atlantic meridional overturning circulation
\parencite{feng_deep_2014}.

It is well known, however, that the dynamics within the two major subcomponents
of the Earth's climate system, ocean and atmosphere, are closely entangled
\parencite{trenberth_decadal_1994,frankignoul_gulf_2001}. Examples for these
interrelationships include the North Atlantic eddy-driven jet stream
\parencite{woollings_variability_2010} or the Pacific ocean forcing to the
atmosphere which is closely related to the dynamics of the El Ni\~no Southern
Oscillation \parencite{wyrtki_nino_1975}. Further, it has been shown that on
time scales of up to one month the ocean is forced by atmospheric circulation, prominently manifested in terms of long-term variability patterns like the Pacific North American pattern
\parencite[e.g.][]{frankignoul_observed_2007} and the North Atlantic
Oscillation \parencite{czaja_influence_1999, gastineau_influence_2015}. 

Inspired by approaches to investigate the interaction structure between
different mutually coupled subsystems such as infrastructure networks
\parencite{vespignani_complex_2010, buldyrev_catastrophic_2010,
  boccaletti_structure_2014} a novel set of coupled network measures has
been proposed by~\textcite{donges_investigating_2011} which provides a general
tool to quantify interdependencies between subcomponents in complex coupled
climate networks. The latter framework has been successfully applied to
investigate interactions between different layers of geopotential height
fields, where each isobaric surface forms a subcomponent of a larger climate
network.  Similarly, coupled climate networks have been constructed to study
ocean-atmosphere interactions in the tropical Pacific
\parencite{feng_three-dimensional_2012} or over the South Atlantic Convergence
Zone \parencite{tirabassi_study_2015}.

Following upon these previous studies, in this work we extend the
approach by~\textcite{donges_investigating_2011} and present an exploratory study
to understand and quantify ocean-atmosphere interactions in the Northern
Hemisphere mid-to-high latitudes during boreal winter at monthly scales. This
temporal restriction is chosen, since previous
studies by means of lagged maximum covariance analysis (MCA) have already
revealed that the statistical interrelationship between atmosphere and ocean is strongest and
most significant at lags of zero or one month during late fall and winter
\parencite[e.g.][]{czaja_influence_1999, wen_observations_2005,
	liu_observational_2006, frankignoul_observed_2007, gastineau_influence_2015}.

To investigate further the spatial structure of these complex interaction
patterns, we construct here in total 18 coupled climate networks consisting of
two layers each, one layer representing sea surface temperature (SST) anomalies
and the other geopotential height fields (HGT) at different pressure levels
from 1000 to 10 mbar covering the entire troposphere as well as the lower
stratosphere.

Our area of study covers the whole Northern Hemisphere north of $30^\circ$N so
that the density of grid points in the considered climate data sets increases
rapidly towards the poles and induces some bias in the unweighted network measures
\parencite{tsonis_what_2006, radebach_disentangling_2013}. Therefore, the standard
coupled network approach by~\textcite{donges_investigating_2011} is not
sufficient in the present case. To overcome the problem associated with the
heterogeneous spatial density of grid points interpreted as nodes of the
climate network,~\textcite{heitzig_node-weighted_2012} introduced a novel set of
network measures that takes into account the different sizes or weights of
nodes in the network. By following an axiomatic approach, each
\textit{standard} (unweighted) network measure can be transformed into its
weighted counterpart, the so-called \textit{node splitting invariant} (n.s.i.)
network measure. Corresponding n.s.i.\ measures have also been derived
by~\textcite{zemp_node-weighted_2014} for edge-weighted and directed networks. 

To quantify the topology of coupled climate networks, we rely in this work on
the previously defined versions of local (i.e.\ node-wise) n.s.i.\ coupled
network measures
\parencite{feng_three-dimensional_2012,wiedermann_node-weighted_2013} and
additionally derive further weighted global network measures following the
approach introduced by~\textcite{heitzig_node-weighted_2012}. This allows us to
assess and compare the macroscopic correlation structure in each of the 18
coupled climate networks.

We compare the results of MCA \parencite[e.g.][]{storch_statistical_2001}, a
well-established standard tool from statistical climatology, with the
cross-degree density of nodes in the different subnetworks and confirm expected
similarities between the two measures \parencite{donges_how_2015}. By utilizing
network measures of higher order such as the n.s.i.\ local cross-clustering
coefficient, we find that the statistical interrelation between ocean and
atmosphere exhibits a hierarchical structure, in which individual parts or
areas of the ocean surface correlate strongly with multiple statistically
dissimilar parts of the atmosphere. Building upon previous studies by, e.g.,
~\cite{frankignoul_observed_2007,czaja_influence_1999}, and
~\cite{gastineau_influence_2015} we relate the observed hierarchy to dominant
atmospheric patterns forcing the ocean on the time scales investigated in this
study.

The remainder of this paper is organized as follows.
Section~\ref{sec:data_methods} introduces the data sets and all methods, i.e.,
maximum covariance analysis and coupled climate network analysis, that are
applied in this study. Section~\ref{sec:results} presents all results of the
analysis followed by conclusions and an outlook discussing future research
tasks in Section~\ref{sec:conclusions}.

\section{Data \& Methods}\label{sec:data_methods}
\subsection{Data description}\label{sec:data_description}
We construct coupled climate networks from two different climatic
observables in order to investigate their interaction structure. One subnetwork
is based on monthly anomalies of geopotential height (HGT) fields
obtained from the ERA40 reanalysis project of the European Centre for
Medium-Range Weather Forecast \parencite{Uppala_ERA-40_2005}. The data is given on
a regular latitude/longitude grid with a spatial resolution of $\Delta\lambda =
\Delta\phi = 2.5^\circ$. In total, we investigate $18$ layers of HGT fields.
The corresponding pressure values at each isobaric surface as well as the average
geopotential height are given in Tab.~\ref{table:HGT}. The second subnetwork is
constructed from the monthly averaged SST field (HadISST1)
provided by the Met Office Hadley Centre \parencite{rayner_global_2003} with a
resolution of $\Delta\lambda = \Delta\phi = 1^\circ$. All grid points with
corresponding time series containing missing values are removed from the data
set as they represent areas which have been at least temporarily covered by sea-ice.

For our analysis we investigate all grid points north of $\lambda=30^\circ$N excluding
the North Pole itself. Both data sets are cropped in their
temporal extent to cover the same time span from January 1958 to December 2001
and, hence, each time series consists of $T=528$ temporal sampling points. We
obtain a total number of $N_s=6201$ grid points for the SST data and $N_i=3456$
grid points for each isobaric surface $i$ of HGT\@. For both data sets, we
remove the annual cycle by subtracting the climatic mean for each month
from each time series. 
Since we focus on the spatial structure of strong statistical
interrelationships between ocean and
atmosphere during boreal winter months (DJF), we use only the
corresponding values which yields a length of each time
series of $\tau = 132$ data points. 

\subsection{Maximum covariance analysis (MCA)}\label{sec:mca}
Consider two sets of time series ${\{X_{s_n}(t)\}}_{n=1}^{N_s}$ and
${\{X_{i_m}(t)\}}_{m=1}^{N_i}$ representing two different climatic
fields, which in the scope of our application are the SST field (in what
follows indicated
by the index \textit{s}) and one layer \textit{i} of HGT (see also
Tab.~\ref{table:HGT}). Further assume each individual time series in both fields to be normalized to
zero mean and unit variance. The
linear lag-zero cross-correlation matrix $\textbf{C}_{si}$
with entries $C_{s_n i_m}$ is then defined as

\begin{align}
C_{s_n i_m} = \frac{1}{\tau} \sum_{t=1}^\tau X_{s_n}(t)
X_{i_m}(t)\label{eqn:cross_covariance},
\end{align}
where $n =1,\ldots,N_s$, $m = 1,\ldots,N_i$ and 
$\tau$ denotes the total number of temporal sampling points in the two time
series. Due to the heterogeneous spatial distribution of
grid points in the present data sets all matrix entries $C_{s_n i_m}$ are additionally multiplied by the square
roots of the cosine of latitudinal positions $\lambda_\bullet$ to ensure equal weighting. This then
yields the weighted cross-correlation matrix $\textbf{C}_{si}^w$ with entries

\begin{align}
C_{s_n i_m}^w = \sqrt{\cos \lambda_{s_n} \cos \lambda_{i_m}} C_{s_n i_m}.\label{eqn:weighted_cross_correlation}
\end{align}
Analogously to empirical orthogonal function (EOF) analysis
\parencite[e.g.][]{ghil_advanced_2002,hannachi_empirical_2007}, MCA
identifies orthonormal pairs of coupled patterns $\vec{p}^{(m)}_s
= (p_{s_1}^{(m)}\ldots
p^{(m)}_{s_{N_s}})$ and $\vec{p}_i^{(m)} = (p^{(m)}_{i_1}\ldots
p^{(m)}_{i_{N_i}})$ for $m=1,\ldots, R$ (with $R$ being the rank of
$\textbf{C}_{si}$) which explain as
much as possible of the covariance between pairs of time series taken
from the two different climatic fields
\parencite[e.g.][]{bretherton_intercomparison_1992,storch_statistical_2001}. The
coupled patterns are obtained by solving the singular value problem of the weighted
cross-covariance matrix,

\begin{align}
	\textbf{C}_{si}^w \vec{p}_i^{(m)} &= \sigma_m
	\vec{p}_s^{(m)},\label{eqn:coupled_patterns1}\\
	{(\textbf{C}_{is}^{w})}^\textbf{T} \vec{p}_s^{(m)} &= \sigma_m \vec{p}_i^{(m)}.
	\label{eqn:coupled_patterns2}
\end{align}
They are ordered according to their respective singular values $\sigma_k$
with $\sigma_1 \geq \sigma_2 \geq \ldots \geq \sigma_R$. Hence,
$\sigma_1$ denotes the largest among the $R$ singular values that can be found to
solve the above equations. Therefore, $\vec{p}_i^{(1)}$ and $\vec{p}_s^{(1)}$ are referred
to as the \textit{leading} coupled patterns representing the largest fraction
of squared covariance between the two climatic fields given by 
$\sigma_1^2$.
\subsection{Coupled climate network construction}
In climate networks, each node represents a climatic time series and
links indicate strong correlations between two series. Hence, the $N \times
N$ ($ N=N_s+N_i$) correlation matrix contains the pairwise linear statistical
relationships between all time series considered for the network construction.
Here, we independently construct coupled climate networks for all combinations of
the SST field and each of the 18 isobaric surfaces of HGT, which shall be
investigated separately and rely on the linear Pearson
correlation coefficient as an appropriate measure of statistical association. Hence, the
correlation matrix has the form

\begin{align}
\textbf{C} = \begin{pmatrix} \textbf{C}_{ss} & \textbf{C}_{si} \\
	\textbf{C}_{is} & \textbf{C}_{ii}\end{pmatrix}.
\end{align}
The two block matrices $\textbf{C}_{ss}$ ($N_s\times N_s$) and $\textbf{C}_{ii}$
($N_i\times N_i$) represent the (internal) correlation matrices of the SST and
HGT fields, respectively, which consist of elements

\begin{align}
C_{s_n s_m} &= \frac{1}{\tau} \sum_{t=1}^\tau X_{s_n}(t) X_{s_m}(t), \quad n,m =
1,\ldots,N_s, \\
C_{i_n i_m} &= \frac{1}{\tau} \sum_{t=1}^\tau X_{i_n}(t) X_{i_m}(t), \quad n,m =
1,\ldots,N_i.
\end{align}
The elements of $\textbf{C}_{si}=\textbf{C}_{is}^\textbf{T}$ are derived according to
Eq.~\eqref{eqn:cross_covariance}.
Note, that (in contrast to the computation of
the leading coupled patterns) we construct the coupled climate networks from the
\textit{unweighted} correlation matrix $\textbf{C}$, while the correction for the
heterogeneous spatial distribution of nodes is implemented into the corresponding network
measures (see Sec.~\ref{sec:interacting_network_characteristics}). 

\begin{figure}[htbp] 
	\centering
	\includegraphics[width=\linewidth]{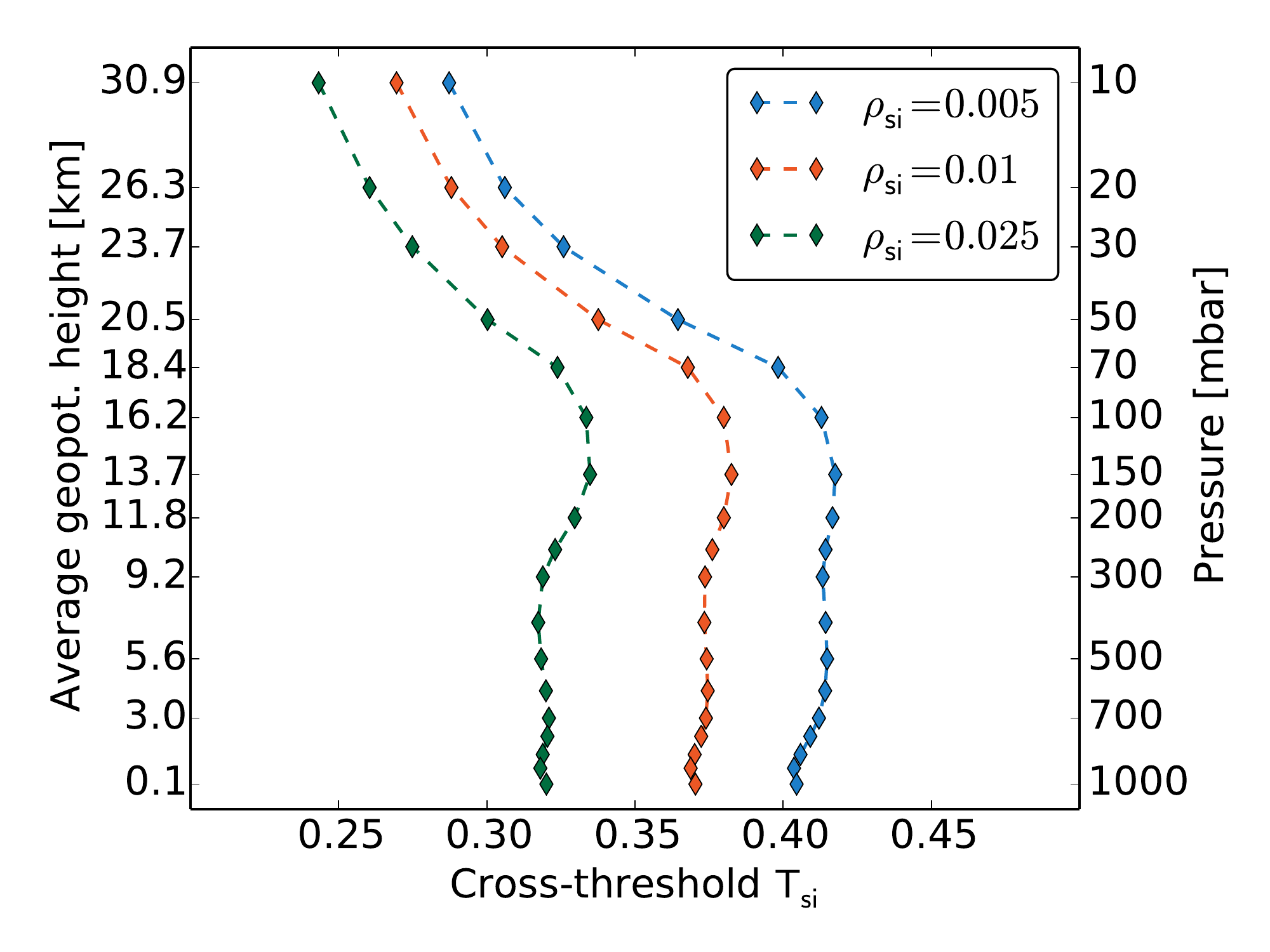}
	\caption{Cross-threshold $T_{si}$ between the subnetwork
		constructed from the SST field and all 18 isobaric surfaces of HGT in
		winter 
		for different standard (unweighted) cross-link densities. 
	}\label{fig:cross-threshold} 
\end{figure}

From the correlation matrix $\textbf{C}$ one generally derives the network's
adjacency matrix $\A^+$ by setting a fixed threshold $T$ such that only a
certain fraction (i.e.\ the link density $\rho$) of strongest correlations is
represented by links in the resulting climate network. For obtaining the
adjacency matrix $\A^+$ of coupled climate networks, we refine this procedure
by fixing a desired link density $\rho_{s}=\rho_{i}=0.01$ for the structure of
internal links within the two subnetworks representing SST and HGT fields,
respectively. This means that only nodes with a correlation above the empirical
99th percentile of correlations between all time series within each field are
connected. This condition then leads to internal correlation thresholds
$T_s=0.8101$ for the SST field and $T_i$ for each isobaric surface of HGT
(Tab.~\ref{table:HGT}). Usually, the dynamics within the different
climatic fields shows much higher cross-correlations than 
between them. We account for this fact by assuming the fraction of significant
interactions between the climatic fields to be lower than those within
them. Specifically, we request a cross-link density of $\rho_{si}=0.005 <
\rho_s = \rho_i$, which is lower than the internal ones, and derive a
cross-threshold $T_{si}$ for each layer of HGT individually
(Fig.~\ref{fig:cross-threshold}). All internal thresholds $T_s$
and $T_i$ are significantly larger than the obtained cross-thresholds $T_{si}$.
Thus, setting a global link density or threshold would cause no or few cross-links to
be present between the two fields or respective subnetworks. We further note
that all links in each of the coupled climate
networks represent correlations that are significant at least at the 95\% confidence
level of a standard \textit{t}-test, where the degrees of freedom are
determined by the total number of temporal sampling points $\tau$ in each of
the time series (with $\tau=132$ we thus obtain 130 degrees of freedom when neglecting the presence of serial correlations in the individual time series).

The different values of $T_{si}$ already give an impression of the 
strength of correlation between the SST field and the different isobaric layers: low
thresholds generally indicate weaker correlations while high thresholds imply
stronger similarity between both fields. Further we note that the resulting
cross-thresholds vary smoothly with the choice of cross-link density
(Fig.~\ref{fig:cross-threshold}) and we thus consider the construction
mechanism to be sufficiently insensitive to the actual choice of cross-link
density. 

Using the different thresholds introduced above, we obtain the coupled
climate network's adjacency matrix by individually thresholding the absolute
correlation values between and within both fields as

\begin{align}
	\nonumber \A^+ &= \begin{pmatrix} \Theta(|\textbf{C}_{ss}|-T_s) 
	& \Theta(|\textbf{C}_{si}|-T_{si}) \\
	\Theta(|\textbf{C}_{is}|-T_{si}) & \Theta(|\textbf{C}_{ii}|-T_i) 
\end{pmatrix} 
,\label{eqn:full_adjacency}
\end{align}
where $\Theta(\cdot)$ denotes the Heaviside function.  Note that in most recent
studies on climate networks self-loops (resulting in a non-vanishing trace of
the adjacency matrix) have been excluded. In this case the adjacency matrix is
usually denoted as $\A$. Since we aim to apply node splitting invariant network
measures (see below) to quantify the network's topology we specifically demand
each node to be connected with itself. The resulting matrix $\A^+$ is referred
to as the \textit{extended} adjacency
matrix~\parencite{heitzig_node-weighted_2012}. Further note, that the usage of the
term \textit{coupled} in \textit{coupled climate networks} does not imply the notion of
any directionality or causal influence between the two fields under study. It
is simply meant to indicate the fact that the network under study is composed
of more than a single climatic field. 

\subsection{Coupled network
	characteristics}\label{sec:interacting_network_characteristics}
The local (point-wise) and global structure of a climate network can be
quantified by a variety of network measures
\parencite{newman_structure_2003,albert_statistical_2002,donges_complex_2009},
which can generally be interpreted as specific operations on the adjacency
matrix. The climate networks in this study are constructed from climate data
sets where the density of grid points and, hence, the density of nodes in the
network, rapidly increases towards the North pole.  In order to avoid a bias in
the evaluation of the climate network's structure, we account for this effect
by relying on node-weighted network measures and value nodes with a gradually
decreasing weight as one moves from the equator towards the pole. To quantify the
correlation structure between ocean and atmosphere
at each node we focus on two previously defined
node weighted local network measures, the n.s.i\
cross-degree~\parencite{feng_three-dimensional_2012} and the n.s.i.\ local
cross-clustering coefficient~\parencite{wiedermann_node-weighted_2013}. In
addition, we utilize the construction mechanism introduced
by~\textcite{heitzig_node-weighted_2012} to convert global interacting network
measures~\parencite{donges_investigating_2011} into their weighted counterparts. 

\subsubsection{Preliminaries}
Consider a coupled climate network $G=(V,E)$ with a set of nodes $V$, links $E$
and the number of nodes $N=|V|$. Following the general naming convention in
the climate network framework we identify every node $v\in V$ with a natural
number $p=1,\ldots,N$, such that $p$ serves as the label of the node as well
as an index to corresponding network characteristics. The network $G$ is represented by its adjacency matrix
$\A$ with $A_{pq} = 1~\text{if}\ (p,q) \in E,\ A_{pq} =  0~\text{if}\ (p,q)
\not\in E$. In this study, each coupled climate network is composed of two subnetworks,
$G_s=(V_s, E_{ss})$ representing the ocean and $G_i=(V_i, E_{ii})$ representing
a specific atmospheric layer. The set of nodes $V$ divides into subsets $V_s$ and $V_i$
such that each node belongs to exactly one subnetwork (i.e. $V=V_s \cup V_i$
and $V_s \cap V_i=\emptyset$). Likewise, the set of links $E$ splits into
internal link sets $E_{ss}$ and $E_{ii}$ (connecting nodes  within a
subnetwork)
and cross-link sets $E_{si}$ connecting nodes $v\in V_s$ with nodes $q\in V_i$
in the subnetworks $G_s$ and $G_i$, respectively~\parencite{donges_investigating_2011}. 

In the present case (as for all regularly gridded climate data sets) the share on the
entire area of the surface that is represented by each node is governed by its
latitudinal position $\lambda_v$ on the grid.
Following~\textcite{tsonis_what_2006}, we therefore assign to each node $v$ in the climate
network a weight

\begin{align}
w_v = \cos\lambda_v.
\end{align}
Note that, by following this convention the climate networks' node weights
$w_v$ exhibit the same dimension as the weights of the cross-covariance matrix in
Eq.~\eqref{eqn:weighted_cross_correlation}.

\textcite{heitzig_node-weighted_2012} introduced a novel set of 
\textit{node splitting invariant} (n.s.i.) network measures to quantify the
topology of a climate network with such a heterogeneous spatial node density for the
case of a single-layer network and, hence, only one climate variable under
study. In fact, the n.s.i.\ network measures are not restricted
to climate networks but can be utilized to study any type of single-layer
complex network where nodes represent entities of different weights.
\textcite{heitzig_node-weighted_2012} 
further showed that each complex network measure can be transformed into its
weighted counterpart by using a four-step construction mechanism:
\begin{enumerate}
\item[(a)] Sum up weights $w_v$ whenever the unweighted measure counts nodes.
\item[(b)] Treat every node $v\in V$ as connected with itself.
\item[(c)] Allow equality in summations over indices $v$ and $q$ wherever the original measure
	involves a sum over distinct nodes $v$ and $q$.
\item[(d)] ``Plug in'' n.s.i.\ versions of measures wherever they are used in
	the definition of other measures.
\end{enumerate}
From the definition of the adjacency matrix $\A^+$ in
Eq.~\eqref{eqn:full_adjacency} we note that step (b) of the above scheme is in
our case already fulfilled.~\textcite{wiedermann_node-weighted_2013}
and~\textcite{zemp_node-weighted_2014} recently utilized the proposed scheme to
convert local interacting network measures as well as measures for directed
networks into their weighted counterparts. Here, we additionally derive
n.s.i.\ versions of some global cross-network measures that were introduced
by~\textcite{donges_investigating_2011}.

\subsubsection{Local measures}
For quantifying local cross-network interactions in coupled climate networks
we rely on two measures, n.s.i.\ cross-degree $k_v^{j*}$ and n.s.i.\
local cross-clustering coefficient $\mathcal{C}_v^{j*}$, that were introduced by
\textcite{wiedermann_node-weighted_2013} and (for the case of the n.s.i.\
cross-degree) by~\textcite{feng_three-dimensional_2012}. These two measures are
defined as

\begin{align}
    k_v^{j*} &= \sum_{q\in V_j} w_q A^+_{vq},\label{eqn:cross_degree} \\
		\mathcal{C}_v^{j*} &= \frac{1}{{(k_v^{j*})}^2} \sum_{p,q\in V_j} A^+_{vp} A^+_{pq} A^+_{qv} w_q w_p\in [0,1].
		\label{eqn:local_clustering}
\end{align}
In contrast to the unweighted cross-degree 

\begin{align}
	k_v^j = \sum_{q\in V_j} A_{vq}^+ \label{eqn:standard_cross_degree}
\end{align}
which simply counts nodes $q\in V_j$ that are connected with $v\in V_i$,
$k_v^{j*}$ is proportional to the share on the considered overall ice-free ocean
or isobaric surface area, respectively, that is
connected with nodes $v\in V_j$ in the other subnetwork. It therefore gives
a notion of how similar the dynamics at a node $v\in V_i$ is to that of the
other climate variable observed at all available grid points.

Similar to $k_v^{j*}$, $\mathcal{C}_v^{j*}$ no longer relies on the counting of distinct
fully connected node triples in the network (as for the classical local clustering
coefficient \parencite{newman_structure_2003}) but on the weighted sum of
occurrences of triples of connected areas within the two subnetworks. It gives
the probability that an area represented by a node $v \in V_i$ is connected
with two mutually connected and, hence, dynamically similar, areas in the opposite
subnetwork. 
In this spirit, $\mathcal{C}_v^{j*}$ estimates how likely areas in two
different climatic fields or subsystems form clusters of statistical equivalence between
them. A local accumulation of such
connected triples represents clusters of closely connected nodes and, in the spirit
of climate networks, strongly correlated regions. 

In order to make the n.s.i.\ cross-degree $k_v^{j*}$ comparable between the two
subnetworks, we normalize it by the maximum possible weight that nodes $v\in
V_i$ can be connected with,

\begin{align}
	\kappa_v^{j*} &= \frac{\sum_{q\in V_j} w_q A^+_{vq}}{W_j}\in [0, 1].
\end{align}
In the spirit of earlier work by~\textcite{donges_analytical_2012} and
\textcite{donner_recurrence_2010}, we refer to this quantity as the \textit{n.s.i.\
	cross-degree density}.  Here, $W_j = \sum_{q\in V_j} w_q$ denotes the total
weight of all nodes $q \in V_j$. For the case of a single-layer network, a measure
similar to the n.s.i.\ cross-degree density has been introduced by
\textcite{tsonis_what_2006} in terms of the \textit{area weighted connectivity},
which quantities the share on the subdomain of interest represented by the entire
network $G$ that is connected with any nodes $v\in V$.

Generally,~\textcite{wiedermann_node-weighted_2013}
and~\textcite{zemp_node-weighted_2014} showed that the weighted local cross-network
measures improve the representation of a network's topology with inhomogeneous
node density within the domain of interest in comparison with its unweighted
counterparts. 

\textcite{donges_how_2015} showed that for the unweighted case cross-degree and
leading coupled patterns display strong similarity if the first coupled
patterns
explain	a high fraction of the system's covariance. A similar assessment can be made
for the similarity between the leading coupled patterns obtained from a
weighted cross-covariance matrix and the n.s.i.\ cross-degree (see supporting
information).

\subsubsection{Global measures}
In addition to local (per node) network measures we also aim to characterize the
macroscopic interaction structure of each pair of coupled climate networks by means of
global network properties. For coupled climate networks a
variety of unweighted measures have been proposed by~\textcite{donges_investigating_2011}.
Here, we utilize the construction mechanism by
\textcite{heitzig_node-weighted_2012} to convert two of them into their
weighted counterparts as well. 

\paragraph{N.s.i.\ global cross-clustering coefficient.}
The global cross-clustering coefficient $\mathcal{C}_{ij}$ of a subnetwork $G_i$ gives
the probability that for a randomly chosen node $v\in V_i$ one finds
neighbors $p,q \in V_j$ that are mutually linked. It is defined as the
arithmetic mean of all local cross-clustering coefficients $\mathcal{C}_v^{j}$, 

\begin{align}
	\mathcal{C}_{ij} = \frac{1}{N_i}\sum_{v\in V_i} \mathcal{C}_v^{j}.
\end{align}
This measure can be converted into its n.s.i.\ counterpart by
calculating the weighted mean of all values of $\mathcal{C}_{v}^{j*}$,

\begin{align}
	\mathcal{C}_{ij}^* = \frac{1}{W_i}\sum_{v\in V_i} w_v \mathcal{C}_v^{j*}.
\label{eqn:global_clustering}
\end{align}
Again, analogously to the interpretation of the local n.s.i.\ measures,
$\mathcal{C}_{ij}^*$ no longer only measures pure node-wise triangular structures but
takes into account the share on the Earth's surface areas involved in the
formation of triangular structures. Generally, large values of
$\mathcal{C}_{ij}^*$ (which
are induced by a dominance of connected triples between the two subnetworks
under consideration) indicate strong transitivity in the underlying
correlation structure.

\paragraph{N.s.i.\ cross-transitivity.} 
The cross-transitivity $\mathcal{T}_{ij}$ gives the probability that two randomly drawn
nodes $p,q \in V_j$ are connected if they have a common neighbor $v \in V_i$.
It is given as

\begin{align}
	\mathcal{T}_{ij} = \frac{\sum_{v\in V_i}\sum_{p\neq q\in V_j} A_{vp}A_{pq}A_{qv}}
	{\sum_{v\in V_i}\sum_{p\neq q\in V_j} A_{vp}A_{qv}}.
\end{align}
Like $\mathcal{C}_{ij}$, the cross-transitivity is a measure of organization 
with respect to the cross-correlation structure in a coupled network
\parencite{donges_investigating_2011}.
However, in contrast to $\mathcal{C}_{ij}$, $\mathcal{T}_{ij}$ takes into account the increasing
influence of nodes with high cross-degree and weighs them more heavily than
nodes with low cross-degree. More importantly it ignores nodes with no links
into the opposite layer, since these nodes display a zero cross-degree. The
node-weighted variant of $\mathcal{T}_{ij}$ can be written as

\begin{align}
	\mathcal{T}_{ij}^* &= \frac{\sum_{v\in V_i}\sum_{p,q\in V_j} w_v A^+_{vp}w_p
		A^+_{pq}w_q A^+_{qv}}{\sum_{v\in V_i}\sum_{p,q\in V_j}w_v A^+_{vp}w_p w_q
		A^+_{qv}}
 =  \frac{\sum_{v\in V_i}w_v{(k_v^{j*})}^2 \mathcal{C}_v^{j*}}{\sum_{v\in
     V_i}w_v{(k_v^{j*})}^2}.\label{eqn:transitivity}
\end{align}
We note that both $\mathcal{C}_{ij}^*$ and $\mathcal{T}_{ij}^*$
similarly evaluate the transitivity of correlations between the two 
climatic variables under study and, hence, quantify a similar network property. They
are derived, however, in a disjoint manner. One measure, $\mathcal{C}_{ij}^*$ is
computed as the weighted average taken over
$\mathcal{C}_{v}^{j*}$. In contrast, despite suggestions
by~\textcite{radebach_disentangling_2013} to decompose the global transitivity into
local contributions, the n.s.i.\ cross-transitivity $\mathcal{T}_{ij}^*$ is
defined solely as a
global network measure with no direct local counterpart. 
It is
important to note that
n.s.i.\
cross-transitivity and n.s.i.\ global cross-clustering coefficient are commonly
asymmetric in the sense that $\mathcal{T}_{ij}^*\neq\mathcal{T}_{ji}^*$ and 
$\mathcal{C}_{ij}^*\neq\mathcal{C}_{ji}^*$.

\section{Results}\label{sec:results}
\subsection{Maximum covariance analysis (MCA)}\label{sec:results_MCA}
\begin{figure}[htbp]
	\centering
	\includegraphics[height=\linewidth]{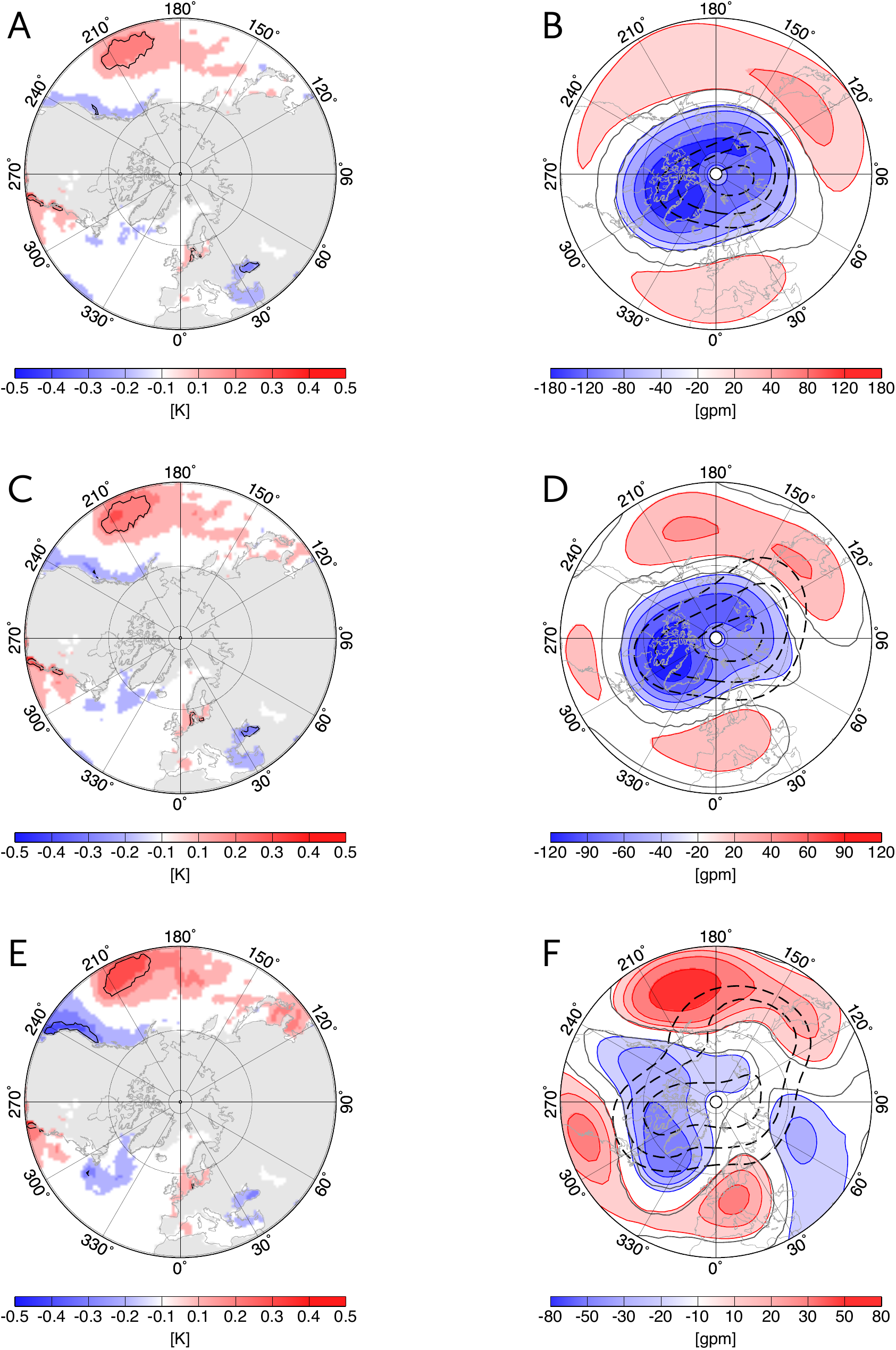}
	\caption{Leading coupled patterns obtained from MCA between the SST field and
		three layers of geopotential height at 50 mbar (A and B), 100 mbar (C and
		D) and 500 mbar (E and F) in winter (DJF). The left column (A, C and E)
		displays the component in the SST and the right column (B, D and F) that in
		the respective HGT field. All spatial patterns are shown as regression maps
		obtained by regressing SST anomalies and geopotential height anomalies onto
		associated time series for the geopotential height field for the respective
		MCA mode. Statistically significant areas at the 95\% confidence level
		based on a two-tailed Student's \textit{t}-test are shown as black contours (for the
		SST maps) and grey contours (for the geopotential height maps). Dashed
    lines indicate the climatic mean geopotential height fields.
	}\label{fig:mca_djf}
\end{figure}

We start our analysis by computing the leading coupled patterns between the SST
field and the 18 HGT layers for boreal winter (DJF). Figure~\ref{fig:mca_djf}
displays the results for three representative layers of HGT at 50 mbar, 100
mbar and 500 mbar.

By applying MCA, we detect coherent large-scale patterns of winter
SST, which co-vary with the winter atmospheric
circulation structures instantaneously. 
The leading MCA patterns explain rather large amounts of 42\%, 63\% and
70\% (for the 500, 100 and 50 mbar pressure level, respectively) of the squared
covariance. At all levels, the leading MCA mode displays
significant SST anomalies over the North Pacific with maximum values
along the sub-Arctic front near 40$^\circ$N, and anomalies of the opposite
sign along the western coast of North America (Fig.~\ref{fig:mca_djf}A,C,E)
\parencite{frankignoul_observed_2007, an_forced_2005}. Over the
North Atlantic, a dipole structure is seen between the northern part
of the Gulf Stream and the Atlantic Ocean south of Greenland including
parts of the Davis Strait and the North Atlantic current. This pattern
resembles the first SST EOF for the Northern Hemisphere during boreal winter (not shown).

This general SST pattern is co-varying with a pressure anomaly pattern showing
a hemispheric annular-like structure in the upper troposphere and
lower stratosphere (Fig.~\ref{fig:mca_djf}B,D). In the mid-troposphere
(Fig.~\ref{fig:mca_djf}F), this pattern displays wave-like deviations from the annular
structure, which show distinct similarities with the wave-train
structure of the Pacific North American (PNA) pattern. Therefore, the leading MCA mode relates negative SST anomalies
along the sub-Arctic front with a positive PNA phase.

The second MCA mode (not shown, explaining 13\%, 17\% and 21\% of the squared
covariance fraction for the 500, 100 and 50 mbar pressure level, respectively)
displays the strongest SST anomalies over the North Atlantic. Over that region,
the SST pattern resembles the northern part of the North Atlantic SST tripole
pattern which is related with the North Atlantic Oscillation (NAO) \parencite[e.g.][]{czaja_influence_1999,
	czaja_observed_2002, gastineau_influence_2015}. Accordingly, the co-varying
atmospheric pattern in the middle troposphere shows the cold ocean/warm land
(COWL) pattern (introduced by~\textcite{wallace_interpretation_1996}) including a
NAO-like dipole over the North Atlantic. At higher
levels, the co-varying atmospheric patterns display a pronounced wave number-2
pattern.

By applying lagged MCA between SST and mid-tropospheric circulation fields,
several studies for the North Atlantic and the North Pacific have shown that
the squared covariance fraction is strongest and most significant at lags of 0 and
1 month during late fall and winter \parencite[e.g.][]{czaja_influence_1999,
	wen_observations_2005, liu_observational_2006, frankignoul_observed_2007,
	gastineau_influence_2015}. This points to the forcing of the SST by the
dominant atmospheric pattern, which is the PNA pattern over the Pacific-North
American sector \parencite[e.g.][]{frankignoul_observed_2007} and the North
Atlantic Oscillation (NAO) over the
North Atlantic-European region \parencite{czaja_influence_1999,
	gastineau_influence_2015}. On the other hand, results of lagged MCA analyses
with the ocean leading by 1 to 4 months in~\textcite{frankignoul_observed_2007}
and~\textcite{gastineau_influence_2015} suggest that the SST anomalies have a
substantial influence on the large-scale atmospheric circulation at these
time-scales.

\subsection{Local coupled network measures}\label{sec:local_measures}
In order to first demonstrate the general consistency of coupled climate network
analysis in comparison with MCA, we continue by generating coupled climate
networks between the SST field and the three previously considered layers of
geopotential height (500 mbar, 100 mbar, 50 mbar).  The n.s.i.\ cross-degree
densities $\ksi$ and $\kis$ are expected to display similar spatial structures
as the corresponding leading coupled patterns \parencite{donges_how_2015} since
the latter explain a high share of the cross-covariance between both fields
(see supporting information).

\begin{figure}[htpb]
	\centering
	\includegraphics[width=\linewidth]{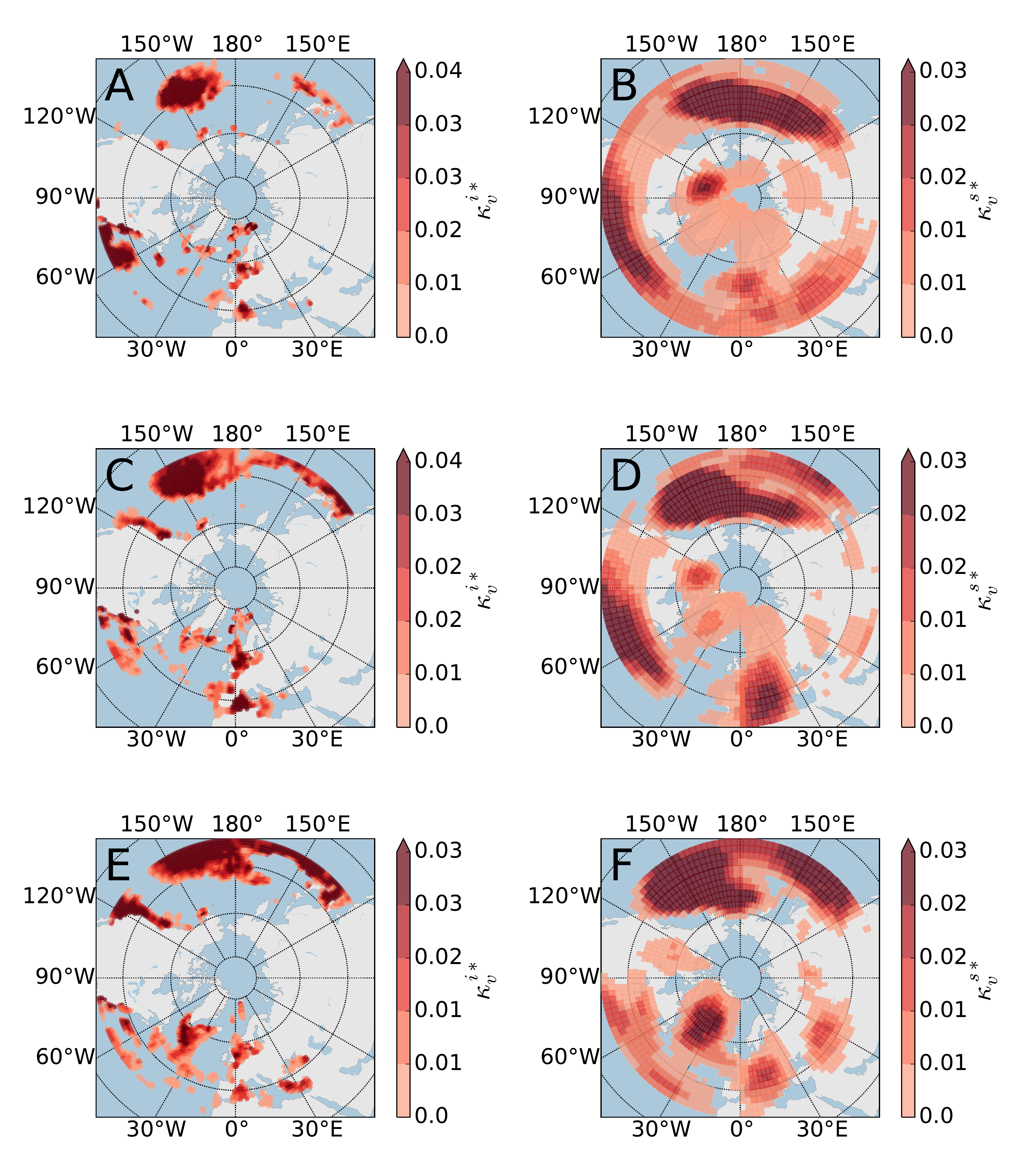}
	\caption{N.s.i.\ cross-degree density for coupled climate networks constructed from
		the SST field and three
		layers of geopotential height at 50 mbar (A and B), 100 mbar (C and D) and
		500 mbar (E and F) for winter months (DJF). The left column (A, C and E) displays
		the n.s.i.\ cross
		degree density $\ksi$ for links pointing from the SST into the HGT subnetwork while 
		the right column (B, D and F) displays the n.s.i.\ cross-degree density $\kis$ for links
		pointing from the HGT into the SST subnetwork. Only nodes with $\ksi>0$ and
		$\kis>0$ are shown.
	}\label{fig:cross_degree_djf}
\end{figure}

As demonstrated in Fig.~\ref{fig:cross_degree_djf}, the results for $\kis$ and
$\ksi$ indeed match well the results obtained from the MCA when comparing the
locations of maximum values in the coupled network's n.s.i.\ cross-degree densities to
those of maximum or minimum values in the leading mode of the MCA\@.  Note,
that the n.s.i.\ cross-degree densities $\kis$ and $\ksi$ take, per definition,
only positive values, while coupled patterns display both, positive and
negative values. Hence, $\kis$ and $\kis$ only reproduce structures that
coincide with the absolute values of the leading coupled patterns.  However, as
only a certain percentage of squared covariance is explained by the leading
coupled patterns, we also note differences between the patterns revealed by the
two methods. In particular, the negative center of action around the North Pole
that is detected by MCA is only weakly present in the cross-degree density fields
$\kis$ for the 50 and 100 mbar HGT fields (compare Fig.~\ref{fig:mca_djf}B,D
with Fig.~\ref{fig:cross_degree_djf}B,D).  For the ocean, preferably the marked
structures in the leading coupled patterns in both the Atlantic and the Pacific
are well recovered by the cross-degree density $\ksi$ while some of the weaker
structures, e.g.\ in the Black Sea, are missing.

Network analysis, however, allows us to undertake a further in-depth analysis
of the correlation structure between the different layers beyond the information provided by MCA. The n.s.i.\ local
cross-clustering coefficients $\csi$ and $\cis$
(Eq.~\eqref{eqn:local_clustering}) give the probabilities that the dynamics at
a grid point in, e.g., the SST field is similar to that at two grid points in
the HGT field, which behave themselves statistically similar. Note that in
the scope of this work we do not account for any possible effects induced by a
common external forcing of the fields, which might artificially induce
correlations and, hence, cause the presence of spurious links between nodes or
triples of nodes. We also do not account for indirect (partial) correlations or
common driver effects within each of the fields when constructing the coupled
climate networks. Conditioning out these possible influences by means of
information theoretic approaches
\parencite{runge_quantifying_2012,runge_quantifying_2013} and causal
effect networks \parencite{kretschmer_causal_2016, runge_identifying_2015} thus remains as a subject of
future research. 

\begin{figure}[htbp]
	\centering
	\includegraphics[width=\linewidth]{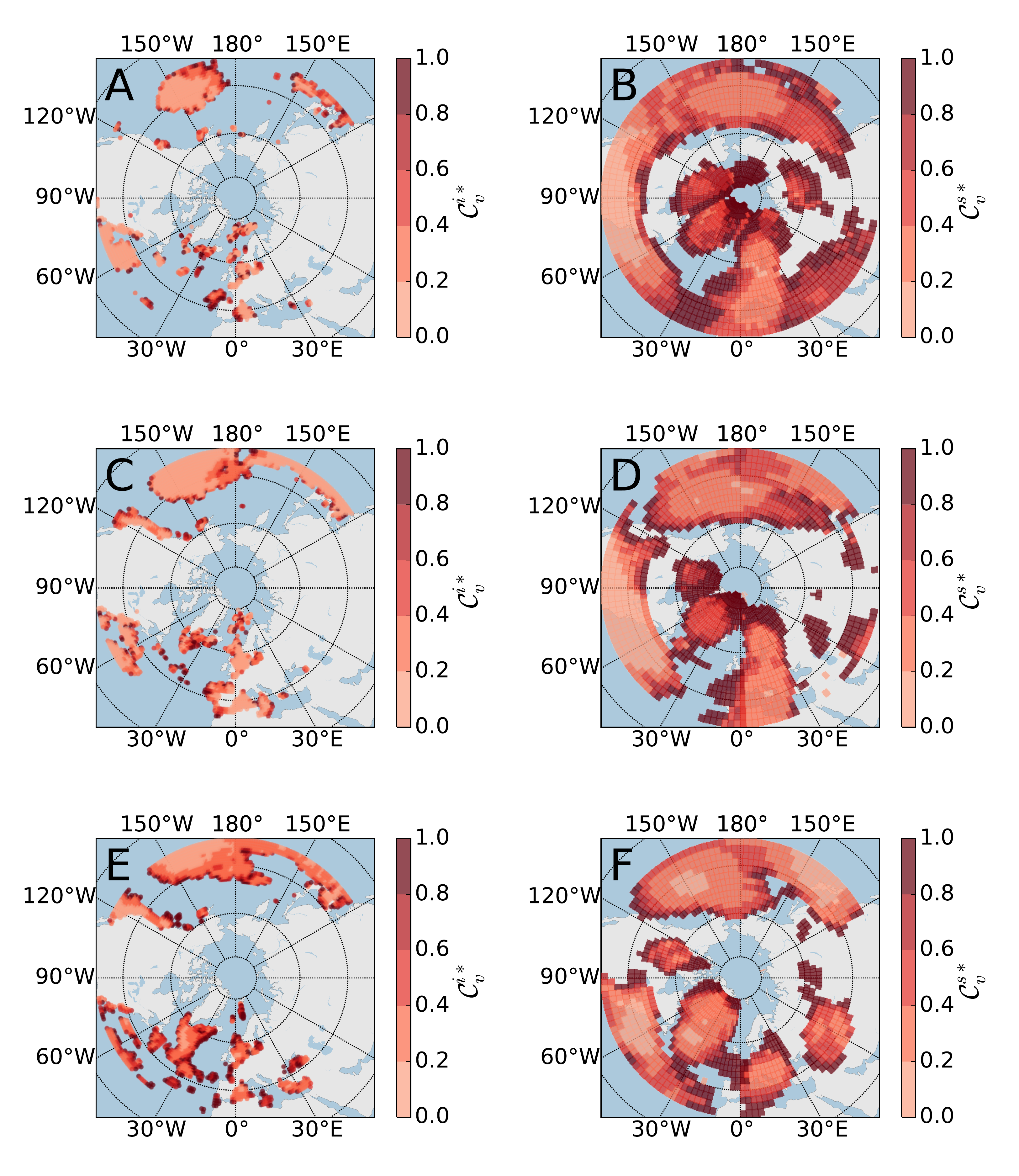}
	\caption{As in Fig.~\ref{fig:cross_degree_djf} for the n.s.i.\ local
		cross-clustering coefficients $\csi$ and $\cis$.}\label{fig:cross_clustering_djf}
\end{figure}

\begin{figure}[htbp]
	\centering
	\includegraphics[width=\linewidth]{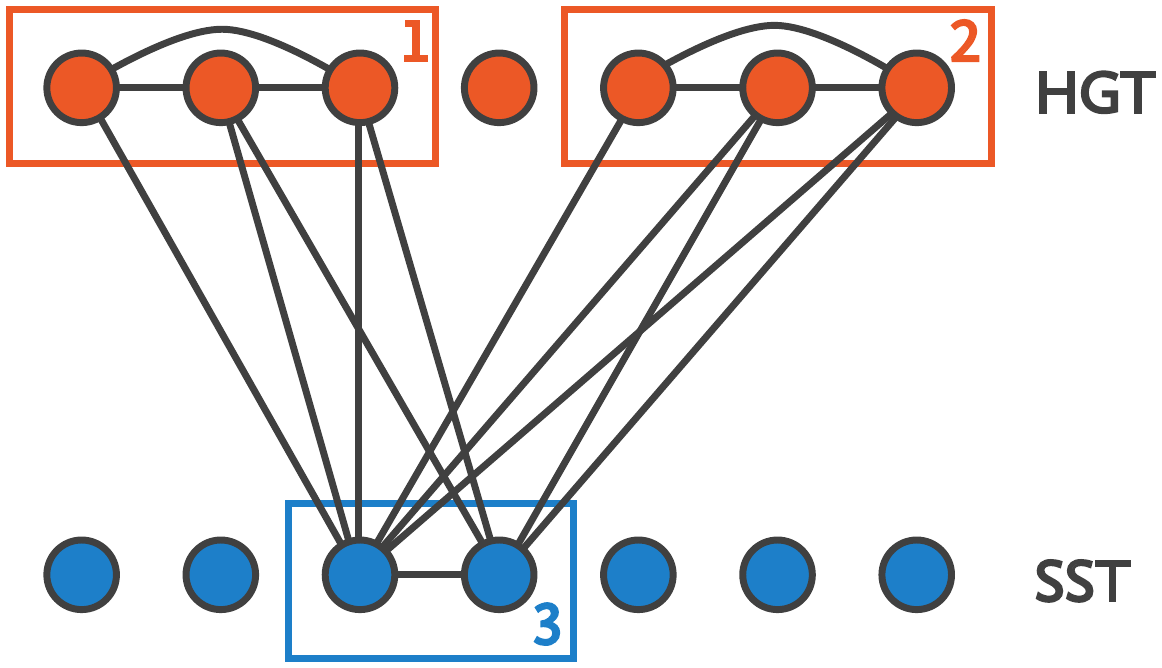}
	\caption{Schematic explanation of the observed quantitative
		differences in the n.s.i.\ local cross-clustering coefficients for nodes in
		the SST and HGT fields. Nodes in the ocean (box 3) tend to connect with
		statistically dissimilar and thus unconnected clusters of nodes in the atmosphere
		(such as nodes in box 1 and 2). Hence, the n.s.i.\ local
		cross-clustering coefficient $\csi$ only takes low values. In contrast, nodes
		in the atmosphere, e.g.\ from box 1, likely connect with 
		clusters in the SST field, such as nodes exclusively in box 3. This
	results in a high n.s.i.\ cross local-clustering coefficient $\cis$ for nodes
in the atmosphere.}\label{fig:interaction_scheme}
\end{figure}

Figure~\ref{fig:cross_clustering_djf} presents the results for the n.s.i.\
local cross-clustering coefficients $\csi$ for nodes in the SST field
(Fig.~\ref{fig:cross_clustering_djf}A,C,E) and $\cis$ for nodes in the HGT
fields (Fig.~\ref{fig:cross_clustering_djf}B,D,F).  Most nodes in the SST field
tend to display a low n.s.i.\ local cross-clustering coefficient $\csi < 0.2$
(Fig.~\ref{fig:cross_clustering_djf}A,C,E) and, thus, preferentially correlate
with nodes in the HGT fields that are mutually dissimilar and therefore
disconnected (Fig.~\ref{fig:interaction_scheme}). In contrast, many nodes in
the HGT fields exhibit a comparatively high or intermediate n.s.i.\ local
cross-clustering coefficient $0.4 < \cis < 1$ (for one of the most prominent examples
compare nodes located at or above the Pacific in
Fig,~\ref{fig:cross_clustering_djf}B,D). Quantitatively, for the
combination of the SST and the 500 mbar HGT field we find an n.s.i.\ global
cross-clustering coefficient (Eq.~\eqref{eqn:global_clustering}) of $\Csi = 0.16$ for SST nodes and $\Cis = 0.28$
for 500 mbar HGT nodes. Ignoring those nodes in the averaging that display zero
n.s.i.\ cross-degree density we obtain values of $\mathcal C_{si}^{*'}=0.42$ and $\mathcal C_{is}^{*'}=0.52$ (note that this definition is different from the one presented in Eq.~\eqref{eqn:global_clustering} as we specifically exclude the contribution of nodes with no links to the opposite subnetwork).
The n.s.i.\ cross-transitivity (Eq.~\eqref{eqn:transitivity}) which weighs nodes according to their n.s.i.\
cross-degree density gives values of $\Tsi=0.2$ and $\Tis=0.25$ for nodes in ocean
and atmosphere, respectively. For all three measures, the values computed for
the atmospheric subnetwork exceed those for the ocean and thus consistently imply that
nodes in the ocean are less likely to connect with mutually connected nodes in
the atmosphere than vice versa. 

\begin{figure}[htbp]
	\centering
  \includegraphics[width=\linewidth]{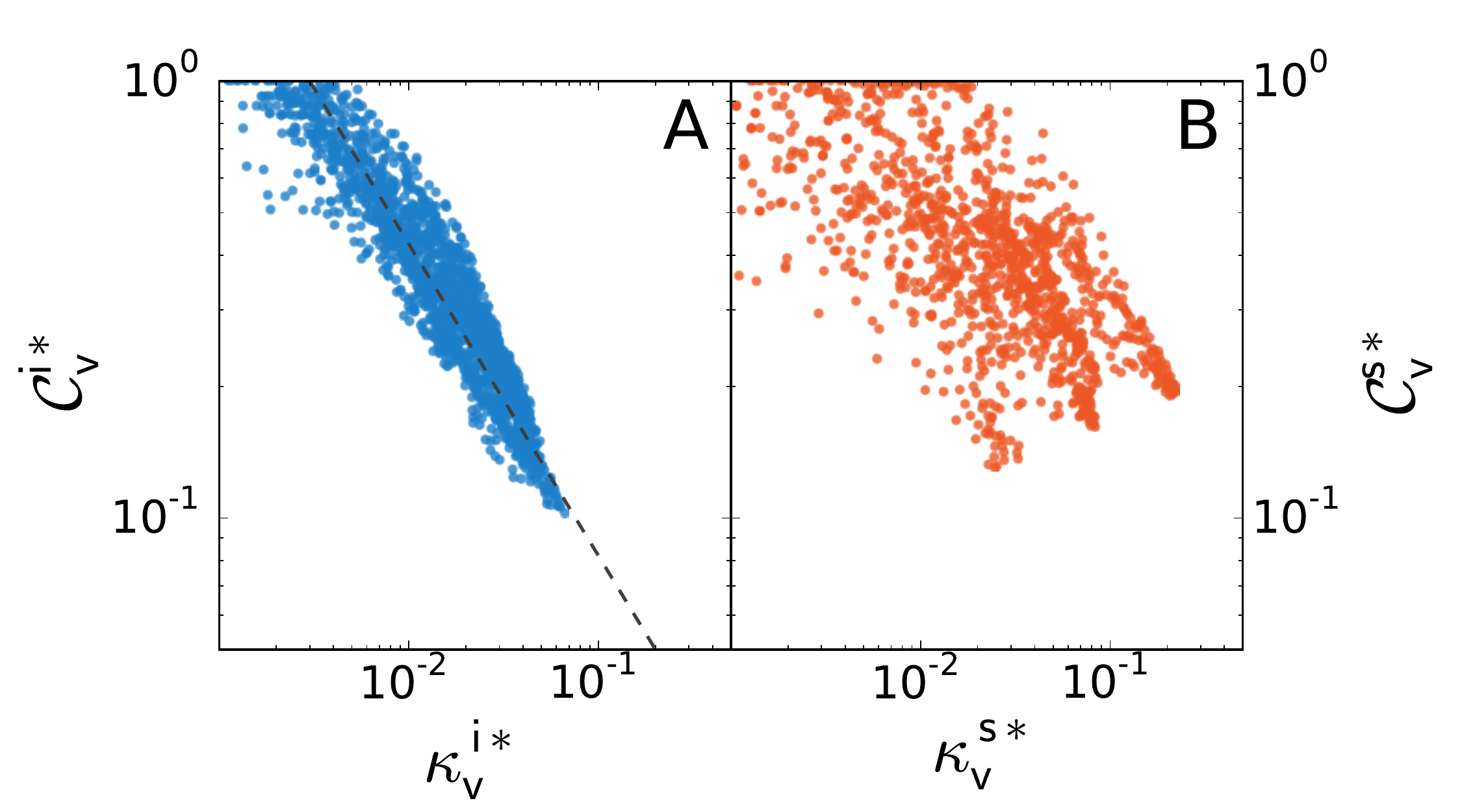}
	\caption{N.s.i.\ local cross-clustering coefficients $\csi(\ksi)$ for nodes
		in the SST field (A) and $\cis(\kis)$ for nodes in the 500 mbar HGT field (B) as
		functions of the respective n.s.i.\ cross-degree densities. The dashed line
		in (A) indicates the relationship $\csi \sim {(\ksi)}^{-\alpha}$ (here
    with $\alpha=0.94$)
		expected for traditional network measures $C_v(k_v)$ in the case of
		hierarchical network structures \parencite{ravasz_hierarchical_2003,ravasz_hierarchical_2002}.}\label{fig:scatter}
\end{figure}

To further quantify the asymmetries in the correlation structure between ocean
and atmosphere, we investigate for each node with a given n.s.i.\ cross-degree
density its corresponding n.s.i.\ local cross-clustering coefficient in a
coupled climate network composed of the SST and 500 mbar HGT fields
(Fig.~\ref{fig:scatter}). This layer is chosen as it provides a good indication
of the atmospheric circulation over the area of
study~\parencite{gastineau_influence_2015,kushnir_atmospheric_2002}. Furthermore,
it displays among the highest values of $T_{si}$ according to
Fig.~\ref{fig:cross-threshold}, which has similarly been described as a
\textit{strong statistical signal} by \textcite{frankignoul_observed_2007}.

For nodes in the SST field (Fig.~\ref{fig:scatter}A), we find that $\csi(\ksi)$
tends to follow a power-law, $\csi\sim {(\ksi)}^{-\alpha}$, which indicates a
hierarchical network structure
\parencite{ravasz_hierarchical_2003,ravasz_hierarchical_2002} which, in contrast, is absent for nodes in the HGT field (Fig.~\ref{fig:scatter}B). Here,
the term \textit{hierarchical} implies that nodes in the SST field strongly correlate with
disconnected clusters of statistically similar nodes in the HGT field as
depicted in Fig.~\ref{fig:interaction_scheme}. This deduction is further
supported by the fact that for the HGT field, the distribution of combinations
of $\cis$ and $\kis$ is more widely spread and $\cis$ generally takes higher
values than $\csi$. This implies that nodes in the HGT field show a stronger
tendency to correlate with mutually connected nodes in the SST field, which can be
assumed to display a strong statistical similarity among themselves
\parencite{molkenthin_networks_2014,tupikina_characterizing_2014}. To test for
the robustness of our results we have
carried out the same analysis as presented in Fig.~\ref{fig:scatter} for
internal link densities of $\rho_{ss}=\rho_{ii}=0.02$ and
$\rho_{ss}=\rho_{ii}=0.05$, and corresponding cross-link densities
$\rho_{si}=0.01$ and $\rho_{si}=0.025$ (see Figs.~1 and 2 in the supporting
information). Even though the power-law exponent $\alpha$ slightly decreases
towards zero with increasing link densities, we find that the qualitative
findings remain unchanged. We thus consider our analysis to be sufficiently
robust with respect to the actual choice of link densities. 

As a remark, we note a general tendency of nodes at the boundaries of a cluster
that links with the opposite field to display comparatively low values of
n.s.i.\ cross-degree density and increased values of n.s.i.\ local
cross-clustering coefficient (Fig.~\ref{fig:cross_degree_djf} and
Fig.~\ref{fig:cross_clustering_djf}). In contrast, nodes located towards the
center of these clusters display increased n.s.i.\ cross-degree density, which
is in general to be expected from the continuity of the underlying system.
However, in that case we also note tendencies for decreased values of n.s.i.\ local
cross-clustering coefficients. This observation is a result of the fact that especially
those nodes with only one link to the opposite field show by definition the
highest value of the n.s.i.\ local cross-clustering coefficient, $\mathcal C_v^{j*}=1$. With
increasing n.s.i.\ cross-degree density this measure converges to a more reasonable
estimate of a node's tendency to cluster. 

One way to address this issue in the future would be to subtract the squared sum
of weights $w_v$ of all neighbors of the considered node from the numerator in
Eq.~\eqref{eqn:local_clustering}. Such procedure would, however, introduce a
non-standard network measure whose properties should be assessed thoroughly in
future research before applying it to climatic studies. To this end, we
acknowledge that the concerned nodes do not play a crucial role for the
propositions put forward in this section, since they (i) are ultimately
dealt with by the assessment of n.s.i.\ cross-transitivity which weighs those
corresponding nodes much lower than those with a high n.s.i.\ cross-degree
density and (ii) only manifest in the very upper left corners of
Fig.~\ref{fig:scatter}A,B. In that case they do not
contribute significantly to the observed relationship between n.s.i.\
cross-degree density and n.s.i.\ local cross-clustering coefficient and have no further
impact on the qualitative statements put forward above. 

\begin{figure}[htbp]
	\centering
  \includegraphics[width=0.7\linewidth]{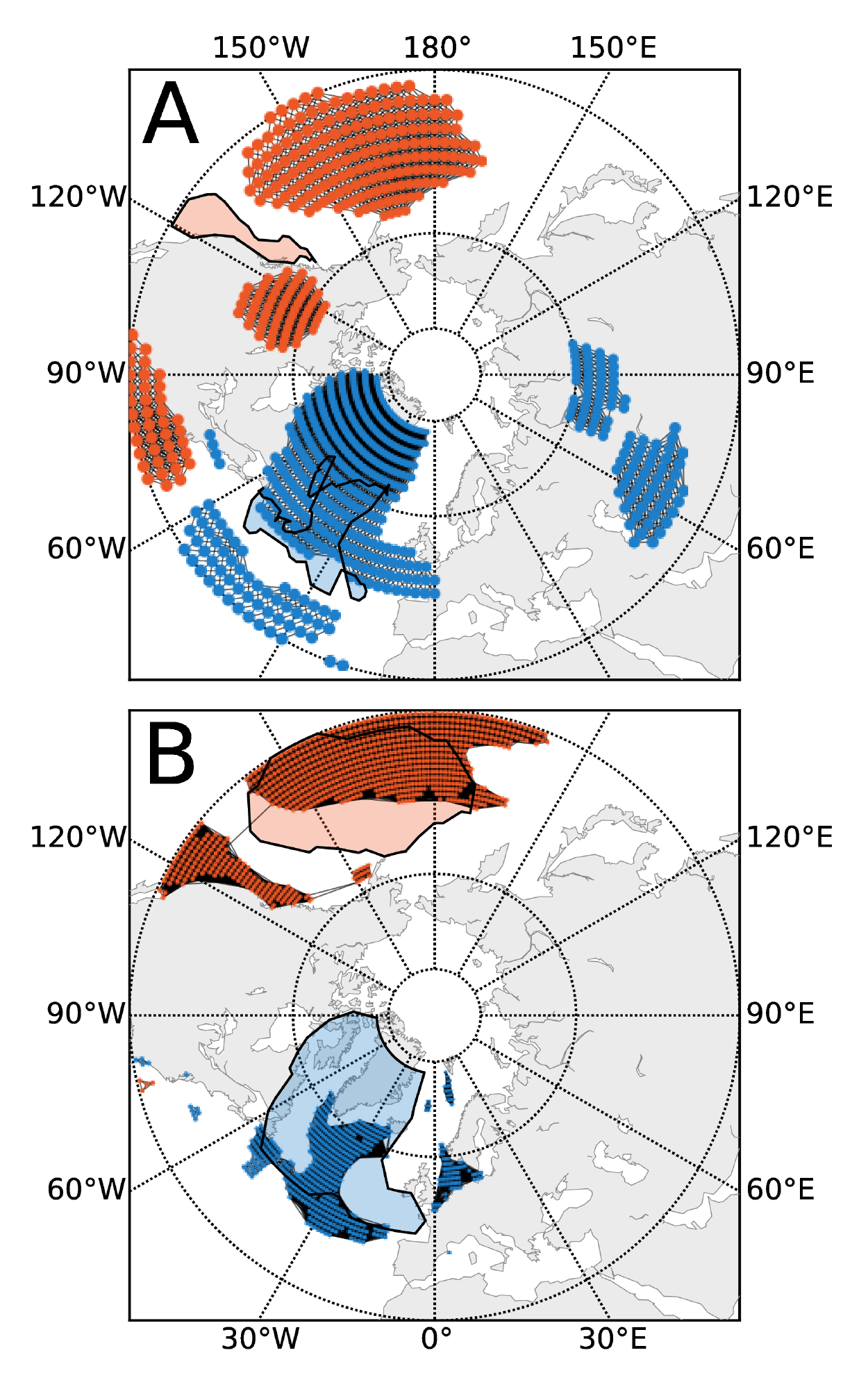}
  \caption{Visualization of a selection of nodes that are relevant for the
    observed hierarchical network structure. (A) Two clusters of nodes in the
    SST field (blue and orange shaded polygons) that show positive values of
    $\ksi$ with the 500 mbar HGT field (compare
    Fig.~\ref{fig:cross_degree_djf}E). Correspondingly, coloured scatter points
    denote nodes in the HGT field that the considered SST nodes are connected
    with. All links that mutually connect the resulting HGT nodes are displayed
    as well.  (B) The same for the two largest patches of nodes in the 500 mbar
    HGT field that where detected in (A). Coloured scatter points now indicate all nodes
    in the SST field that are connected with these
    patches.}\label{fig:hierarchy}
\end{figure}

Following upon the quantitatively observed hierarchy, Fig.~\ref{fig:hierarchy}
allows for a visual inspection of some illustrative parts of the corresponding
network structure. In particular, we display for two selected patches of nodes
in the SST field that show high values of $\ksi$ with the HGT field (blue and
orange shaded polygons in Fig.~\ref{fig:hierarchy}A) their corresponding
neighboring nodes in the HGT field as well as all links between those nodes
(respective blue and orange scatter in Fig.~\ref{fig:hierarchy}A). While
ignoring very small clusters we find in total four (three) substantial mutually
disconnected patches of nodes in the HGT field that correlate with the
respective ocean patches.  Vice versa, by selecting the resulting two largest
patches of nodes in the HGT field above both oceans (blue and orange shaded
polygons in Fig.~\ref{fig:hierarchy}B) we find that each of the patches only
correlates with two disconnected patches of nodes in the SST field that are of
a relevant spatial extent to have an effect on the estimation of $\cis$. Thus,
the resulting n.s.i.\ cross-clustering coefficient $\csi$ for nodes in the
ocean exceed $\cis$ for nodes in the atmosphere as the ocean correlates with
more mutually disconnected clusters of nodes than vice versa. 

Comparing the observed node patches in the HGT field with atmospheric patterns
of large-scale variability patterns ~\parencite{TELLUSA19777}, we relate the
two atmospheric clusters in the HGT field that are located above the Atlantic
(blue scatter in Fig.~\ref{fig:hierarchy}A) with the NAO.  Correspondingly, the
three patches located above the Pacific (orange scatter in
Fig.~\ref{fig:hierarchy}A) coincide well with the spatial signature of the PNA pattern. Taking into
account past studies that applied lagged correlation analysis we note that on
the time scales considered in this study the atmosphere serves as a driving
force of the ocean along the spatial domain that is of interest here
\parencite[e.g.][]{frankignoul_observed_2007,czaja_influence_1999,
  gastineau_influence_2015}. Thus, the hierarchical network structure might on
the one hand be a result of the aforementioned atmospheric forcing to
the ocean. On the other hand, with reference to Fig.~\ref{fig:hierarchy},
the framework of coupled climate networks and the methodology put forward in
this work serve to resolve the corresponding induced correlation structure
between the two climatic subsystems in a spatially explicit way, such that it
enables to specifically detect forcing and forced areas in atmosphere and ocean,
respectively. 

Choosing different HGT layers up to 200 mbar yield similar results (not shown).
This aligns well with previous results by~\textcite{czaja_influence_1999}
and~\textcite{frankignoul_observed_2007} who observed comparable spatial
statistical patterns at each tropospheric level. Thus, the observed hierarchy
seems to be a generic property of the troposphere. Following the above lines
of thought, future work should investigate coupled climate networks constructed
from lagged cross-correlations to investigate whether the observed structures
are indeed a result of short-term atmosphere-to-ocean forcing. Such procedure
would, however, require the derivation of novel directed interacting network
measures, which in turn would provide a valuable extension to the framework of
climate network analysis.

\subsection{Global measures}\label{sec:global_measures}
So far we have focused our study on three atmospheric layers, namely the 50
mbar, 100 mbar and 500 mbar HGT field. Specifically for the latter case, we
have carried out a further in-depth analysis of the observed hierarchical
structures by means of assessing the power-law relationship between $\ksi$ and
$\csi$ as well as investigating the distinct spatial distribution of nodes and
links in ocean and atmosphere that obey the observed hierarchical organization
(Fig.~\ref{fig:hierarchy}). To show that these structures are (i) not only
present for the 500 mbar field and (ii) their observations are robust with respect to the 
choice of link densities we investigate global network characteristics that
provide a macroscopic description of the observed network structures.

Specifically, we study the n.s.i.\ cross-transitivity $\Tsi$ computed
over nodes in the SST field and $\Tis$ computed over nodes in each of the HGT
fields according to Eq.~\eqref{eqn:transitivity} together with the n.s.i.\
global cross-clustering coefficients $\Csi$ and $\Cis$, respectively
(Eq.~\eqref{eqn:global_clustering}). Note again that the latter are defined as
the weighted means of their local counterparts that are presented in
Fig.~\ref{fig:cross_clustering_djf}, where nodes with no links to the opposite
field are weighted in the same fashion as those with adjacent cross-links. In
contrast, the n.s.i.\ cross-transitivity assigns nodes a weight corresponding
to their n.s.i.\ cross-degree (which for the case of no adjacent cross-links
takes a value of zero) and, thus, excludes them from the averaging. 

\begin{figure}[htbp]
	\centering
  \includegraphics[width=\linewidth]{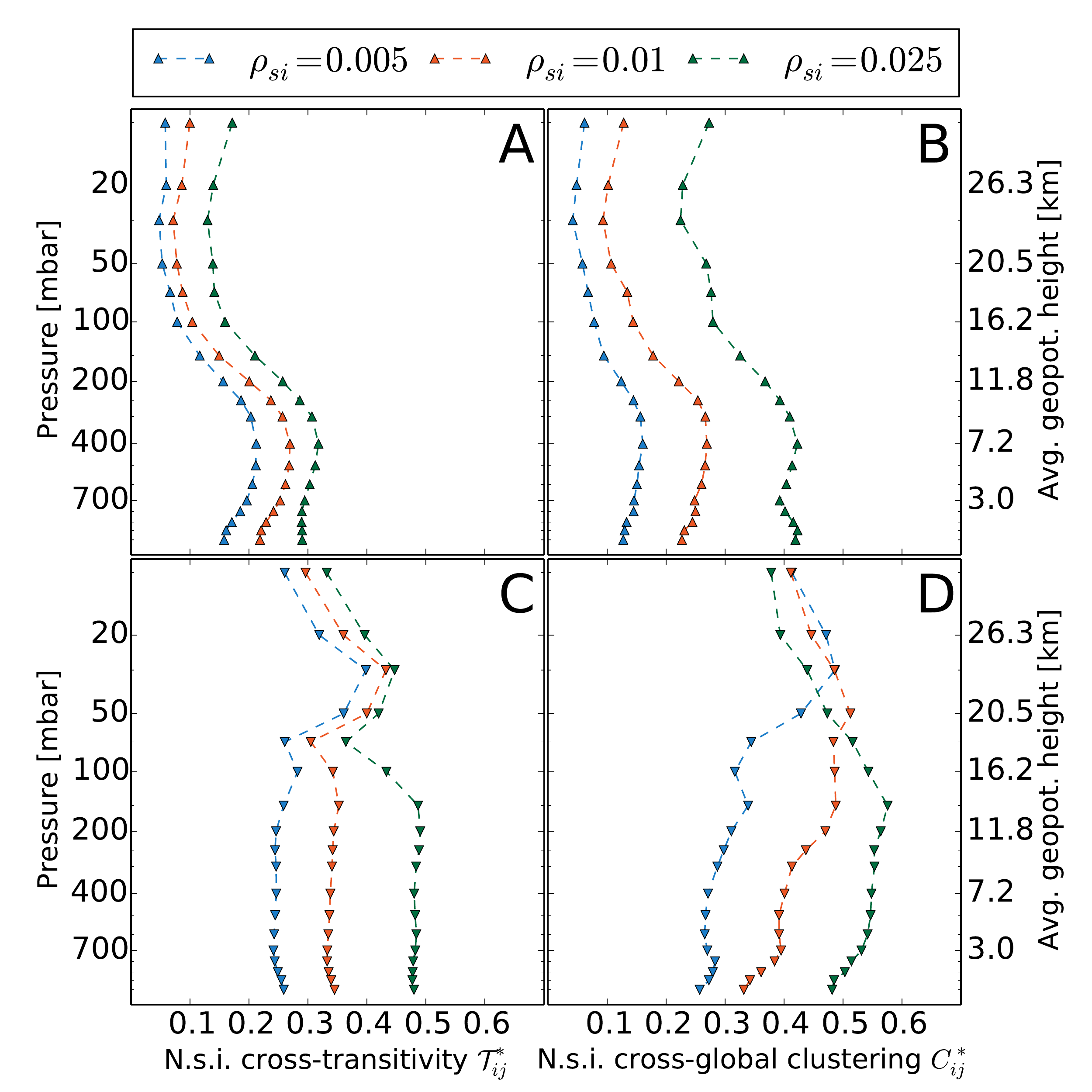}
	\caption{Global coupled network measures computed for all 18 coupled
		climate networks: (A) N.s.i.\ cross-transitivity and (B) n.s.i.\ global
		cross-clustering coefficient (B) taken over all nodes in the SST field. (C)
		and (D) display the respective measures computed over all nodes in the HGT
		field. To demonstrate the robustness and consistency of the results, we
		construct the networks for different choices of (unweighted) cross-link
		density $\rho_{si}$ and internal link density $\rho_i = \rho_s =
		2\rho_{si}$.}\label{fig:global_measures_djf}
\end{figure}

The corresponding results are summarized in Fig.~\ref{fig:global_measures_djf}.
We find that both $\Tsi$ and $\Csi$ show their maximum values at around 10 km
altitude (250 mbar)
(Figs.~\ref{fig:global_measures_djf}A and~\ref{fig:global_measures_djf}B).  For
the same quantities, distinct minima at 850 mbar (1.4 km) coincide with the
transition from the atmospheric boundary layer to the lower troposphere as also
found in~\textcite{donges_investigating_2011}. For all layers above 100 mbar,
$\Tsi$ remains almost constant at low values. Hence, $\Tsi$ and $\Csi$ seem to
naturally discriminate between three different atmospheric layers: below 850
mbar (atmospheric boundary layer), between 850 mbar and 100 mbar (free
troposphere) and above 100 mbar (lower stratosphere).

For the global measures computed over all nodes in the HGT field, we find that
the n.s.i.\ cross-transitivity $\Tis$ shows almost constant values for all
layers below 200 mbar and, hence, again separates well the dynamics within the
troposphere from that inside the stratosphere
(Fig.~\ref{fig:global_measures_djf}C). For all layers above 200 mbar $\Tis$
becomes almost independent of the cross-link density $\rho_{si}$ that is fixed
when constructing the network. The same property also holds for the n.s.i.\
global cross-clustering coefficient $\Cis$ computed over all nodes in the different HGT
fields (Fig.~\ref{fig:global_measures_djf}D).

In agreement with the local measures discussed in Sec.~\ref{sec:local_measures} 
we find that n.s.i.\ cross-transitivity and n.s.i.\ global cross-clustering coefficients
are in most cases larger for nodes in the HGT fields than for nodes in the SST field (compare
Fig.~\ref{fig:global_measures_djf}A with Fig.~\ref{fig:global_measures_djf}C
and Fig.~\ref{fig:global_measures_djf}B with
Fig.~\ref{fig:global_measures_djf}D). As for the n.s.i.\ local cross-clustering
coefficients this indicates again the hierarchical network structure, i.e., a
tendency for nodes in the HGT field to form triangular structures with
nodes in the SST field, that is present across all atmospheric layers ranging
from the troposphere to the lower stratosphere. The
detailed structure of this hierarchy, however, seems to vary with
the different atmospheric layers under study.

This observation further holds not only for the case of $\rho_{si}=0.005$ that
was used in the previous sections but also for larger values that are chosen
from a reasonable range (Fig.~\ref{fig:global_measures_djf}). Thus, we consider our results to be sufficiently
robust with respect to the choice of the networks' link densities.

In general, we observe that the quantitative and qualitative properties of the
n.s.i.\ cross-transitivity and n.s.i.\ global cross-clustering coefficients
vary with the different atmospheric layers. Hence, these global
characteristics may serve to inter-compare and distinguish
between different correlation structures in a coupled climate network.
An in-depth analysis of the mechanisms that cause the occurrence of this
behavior in our specific application remains as a subject of future research.

\section{Conclusions \& Outlook}\label{sec:conclusions}
We have carried out a detailed analysis of the correlation structure between
atmospheric and ocean dynamics in the Northern Hemisphere extratropics from the viewpoint of
coupled climate networks. Comparison between the n.s.i.\ cross-degree density
(measuring the weighted share of significant correlations between grid points
in different layers) and the leading mode of the maximum covariance analysis
(MCA) reveals an expected high congruence between both methods for the
considered data sets. However, coupled network analysis, and particularly the
investigation of higher-order network parameters, allows us to further
disentangle the underlying correlation structure. The (average) n.s.i.\
cross-degree density in combination with the (average) n.s.i.\ local
cross-clustering coefficient provides additional insights on areas in the ocean and the
atmosphere that show strong mutual correlations as well as localized versus
delocalized correlation structures with the respective opposite field. In the
SST field nodes tend to correlate with multiple mutually unconnected groups of
similar nodes within the respective HGT fields. From investigating the
interdependency between n.s.i.\ cross-degree density and n.s.i.\ local
cross-clustering coefficient, we have found that the correlations between the
ocean and the atmosphere exhibit a hierarchical structure in the sense of a power-law relationship between both properties. A visual
inspection of the coupled climate network for the case of the 500 mbar HGT
field reveals that the observed structure could be related with a forcing of the
ocean by the dominant atmospheric patterns above the Atlantic and the Pacific.
Ultimately, global network characteristics further support the results obtained from their
local correspondents by showing that the observed structure is valid for large
parts of the atmosphere ranging from the troposphere to the lower
stratosphere. 

In order to discriminate between the internal variability of the fields under study and
possible influences of an external forcing, future work should analyze ensemble simulations of general
circulation models to rule out common driver effects or assess the
likelihood of their influence on the observed structures. In order to
investigate the influence of spatio-temporal auto-correlation on the outcome
of the present analysis the network could be alternatively constructed
by estimating pairwise thresholds from surrogate data as proposed by
~\cite{palus_discerning_2011}. This approach would, however, break the
comparability of the network approach with that of maximum covariance
analysis, such that a different way of validating and comparing the results
must be found. Comparability could be achieved by assessing synthetic model
data, e.g., created from an auto-regressive process based on principal
components of the data sets under study, and the application of both, MCA and
network analysis. In addition to probable influences of auto-correlation such a
process would allow to assess the influence of different time scales in ocean
and atmosphere on the involved network characteristics.  

Besides all possible
future lines of work with respect to the climatic side of this work, from
a network-theoretic point of view it is worthwhile to construct climate networks
using more advanced causal estimators
\parencite[e.g.][]{runge_quantifying_2012,runge_quantifying_2013,runge_identifying_2015}
to disentangle direct from indirect or externally and internally forced
correlations. 

To this end, our analysis has only been performed for the pairwise correlation
between one atmospheric layer and the ocean. Future studies should further
explore the possibility to refine the proposed methods to also quantify
interactions in a climate network existing of more than two
subnetworks. Specifically, when studying coupled climate networks in the
Northern Hemisphere, one should also consider Arctic sea ice as
an additional observable in the network construction. Its dynamics has already
been discovered as an influencing factor on atmospheric teleconnections and the
dynamics of land snow cover in the Northern Hemisphere
\parencite{handorf_impacts_2015}. The study of coupled climate networks can help
here to further disentangle and quantify possible changes in correlations
between ocean and atmosphere over the course of the past decades that may have
been induced by processes related to the Arctic amplification
\parencite{serreze_arctic_2006}.  Moreover, it is of great interest to apply our
methods not only to coupled networks composed of different climatic
fields (as presented in this work), but also to networks constructed
from just one single climatic field that divides into dynamically
distinct areas \parencite{hlinka_regional_2014} or communities
\parencite{tsonis_community_2010,steinhaeuser_multivariate_2011}. The framework
presented in this work could then be utilized to study and quantify
correlations between these detected or defined regions on or parallel to the
Earth's surface. This would allow for a detailed investigation of correlation
structures between different climatic subsystems such as, for example, the
Indian Summer Monsoon and the El Ni\~no Southern Oscillation.

Finally, it remains to remark that the weighted network measures presented in
this work provide a general framework which can be applied to quantify
interdependencies in complex networks representing subjects of study taken from
many other fields beyond climatology.

\section*{Acknowledgements}
	MW and RVD have been supported by the German Federal Ministry for Education
	and Research via the BMBF Young Investigators Group CoSy-CC$^2$ (grant no.
  01LN1306A) and the Belmont Forum/JPI Climate project GOTHAM. JFD is grateful for financial support by the Stordalen
	Foundation (via the Planetary Boundary Research Network PB.net) and the Earth
	League's EarthDoc program.  JK acknowledges the IRTG 1740 funded by DFG and
	FAPESP\@.  Coupled climate network analysis has been performed using the
	Python package \texttt{pyunicorn}~\parencite{donges_unified_2015} that is
	available at \url{https://github.com/pik-copan/pyunicorn}.
%\end{acknowledgements}

%\bibliographystyle{agsm}
%\bibliographystyle{plainnat}

\printbibliography

\clearpage

\clearpage
\clearpage

\clearpage

\clearpage

\clearpage

\clearpage

\clearpage

%\begin{figure}[htbp]
%	\centering
%	\includegraphics[width=\linewidth]{wiedermann_figure7.pdf}
%	\caption{Zonal averages of the n.s.i.\ cross-degree density (A) $\zonksi$ for
%		nodes in the SST field and (B) $\zonkis$ for nodes in the HGT fields and the n.s.i.\ local
%		cross-clustering coefficients (C) $\zoncsi$ and (D) $\zoncis$. For the Pacific, all grid
%		points between $\phi= 160^\circ$E and $\phi=140^\circ$W and for the Atlantic all grid
%		points between $\phi= 60^\circ$W and $\phi=0^\circ$ longitude are zonally
%		averaged. Areas with no data or average n.s.i\ cross-degree density
%		$\zonksi= 0$ and $\zonkis = 0$ are
%		displayed in grey.
%	}\label{fig:avg_DJF}
%\end{figure}
%\clearpage
%
%\begin{figure}[htbp]
%		\centering
%		\includegraphics[width=\linewidth]{wiedermann_figure8.pdf}
%		\caption{Average zonal wind speed over the Pacific taken over all grid points
%			between $\phi=150^\circ$W and $\phi=120^\circ$E.}\label{fig:zonal_wind}
%\end{figure}
%\clearpage

%\begin{figure}[htbp]
%	\centering
%	\includegraphics[width=0.5\linewidth]{wiedermann_figure9.pdf}
%	\caption{(A) N.s.i.\ cross-link density $\rho_{si}^*$ between the SST field
%		and all 18 layers of geopotential height (HGT) together with the area-weighted average
%		cross-degree densities (B) $\meanksi$ for nodes in the SST field and (C)
%		$\meankis$ for nodes in the HGT
%		fields as defined in Eq.~\eqref{eqn:meanksi} and~\eqref{eqn:meankis}.\label{fig:cross_ld_djf}}
%\end{figure}
%
%\clearpage

\clearpage

\begin{table}[htbp]
		\caption{Air pressure $p_i$ and associated mean geopotential height $Z_i$
			as well as the internal threshold $T_i(\rho_{ii}=0.01)$ corresponding to 
			an internal link density of $\rho_{ii}=0.01$ for each isobaric
			surface $i$.}\label{table:HGT}	
    \centering
    \begin{tabular}{cccc}%\toprule
		\begin{tabular}[c]{@{}c@{}}Layer \\ i\end{tabular} &	
		\begin{tabular}[c]{@{}c@{}}Air pressure\\ $p_i$ [mbar]\end{tabular} &	
		\begin{tabular}[c]{@{}c@{}}Geopotential\\ height $Z_i$ [km]\end{tabular} &
		\begin{tabular}[c]{@{}c@{}}Threshold \\ $T_i(\rho_{ii}=0.01)$\end{tabular} 
		\\\hline
		0  & 10   & 30.9 & 0.9919  \\
    1  & 20   & 26.3 & 0.9936 \\
    2  & 30   & 23.7 & 0.9932 \\
    3  & 50   & 20.5 & 0.9876 \\
    4  & 70   & 18.4 & 0.9781 \\
    5  & 100  & 16.2 & 0.9621\\
    6  & 150  & 13.7 & 0.9263\\
    7  & 200  & 11.8 & 0.9166 \\
    8  & 250  & 10.4 & 0.8982 \\
    9  & 300  & 9.2  & 0.8894\\
    10 & 400  & 7.2  & 0.8895\\
    11 & 500  & 5.6  & 0.8958\\
    12 & 600  & 4.2  & 0.9036\\
    13 & 700  & 3.0  & 0.9119\\
    14 & 775  & 2.2  & 0.9171\\
    15 & 850  & 1.4  & 0.9205\\
    16 & 925  & 0.8  & 0.9215\\    
		17 & 1000 & 0.1  & 0.9197 \\%\bottomrule
	\end{tabular}%\centering
\end{table}

\end{document}